\documentclass[draftcls, onecolumn, 10pt]{IEEEtran}
\usepackage{amssymb,amsmath}
\usepackage[dvips]{graphicx}
\usepackage{amsfonts}
\usepackage{subfigure}
\usepackage{booktabs,multirow}
\usepackage{threeparttable}
\usepackage{color}
\usepackage{cite}
\usepackage[table]{xcolor}

\IEEEoverridecommandlockouts

\begin{document}

%\title{Cognitive Spectrum Sharing With Bi-directional Secondary System}
\title{Distributed and Multi-layer UAV Network
for the Next-generation Wireless Communication
}

\author{\normalsize
Yiming Huo$^1$(ymhuo@uvic.ca), Xiaodai Dong$^1$(xdong@ece.uvic.ca), Tao Lu$^1$(taolu@uvic.ca), \\Wei Xu$^1$\textsuperscript{,}$^2$(wxu@seu.edu.cn), Marvin Yuen$^3$(marvinyu@usc.edu),\\
\vspace{0.70cm}
\small{
$^1$ Department of Electrical and Computer Engineering\\
University of Victoria, BC V8P 5C2, Canada\\
%Email: \{ymhuo}@uvic.ca\\
\vspace{0.2cm}
$^2$	National Mobile Communications Research Laboratory\\
Southeast University, Nanjing 210096, China\\
\vspace{0.2cm}
$^3$ Viterbi School of Engineering, \\
University of Southern California, Los Angeles, California 90089, U.S.A \\

}
\thanks{\textbf{This work has been submitted to the IEEE for possible publication. Copyright may be transferred without notice, after which this version may no longer be accessible.}}
}

%\date{\today}
\renewcommand{\baselinestretch}{1.2}
\thispagestyle{empty}
\maketitle
\thispagestyle{empty}
%\newpage
\setcounter{page}{1}\begin{abstract}
Unmanned aerial vehicles (UAVs) for wireless communications has rapidly grown into a research hotspot as the mass production of high-performance, low-cost, intelligent UAVs become more practical and feasible.  In the meantime, fifth generation (5G) wireless communications is being standardized and planned for deployment globally. During this process, UAVs are gradually being considered as an important part of 5G and expected to play a critical role in enabling more functional diversity for 5G communications. In this article, we conduct an in-depth investigation of mainstream UAV designs and state-of-the-art UAV enabled wireless communication systems. We propose a hierarchical architecture of UAVs with multi-layer and distributed features to facilitate a smooth integration of different mainstream UAVs into the next-generation wireless communication networks. Furthermore, we unveil the critical comprehensive design tradeoffs, in light of both communication and aerodynamic principles. Empirical models and satellite measurement data are used to conduct numerical analysis of the meteorological impacts of UAV enabled, 5G high bands communications.     

\end{abstract}

\IEEEpeerreviewmaketitle

\newpage
\section{Introduction}

In recent years, unmanned aerial vehicles (UAVs) have experienced a rapid transition from the initial military exploitation and aviation industry, to current fast-growing civilian applications such as industrial inspection, scientific research, agricultural practice, security surveillance, emergency rescue, entertainment, etc. In the meantime, the fifth generation (5G) wireless network is being planned for rapid deployment, and many research and industry communities have been seeking diverse paradigms to accelerate this progress and enrich application scenarios. UAV-aided 5G wireless has sparked a large interest as it can facilitate various use cases such as those speculated in the three key principle application scenarios of the International Telecommunication Union (ITU) \cite{ITU2015}. They are namely enhanced mobile broadband (eMBB), ultra reliable low latency communications (uRLLC), and massive machine type communications (mMTC). For example, UAV can play a critical role in providing network service recovery in a disaster-stricken region, enhancing public safety networks, or handling other emergency situations when uRLLC is required. In particular, UAV-assisted eMBB can be considered as an important complement to the 5G cellular network where a 1000 times comprehensive performance improvement over 4G is expected. 

Since July 2016, when the Federal Communication Committee (FCC) adopted a new Upper Microwave Flexible Use Service\footnote{\noindent https://apps.fcc.gov/edocs\_public /attachmatch/FCC-16-89A1.pdf [Accessed: 31-Mar-2018]}, millimeter wave (mmWave) bands for cellular services has become an immediate reality. However, alongside promising opportunities, e.g., larger bandwidths and faster speed, mmWave cellular communications face significant challenges, particularly for terrestrial environments which experience large propagation loss and shadowing effects. The propagation loss challenge can be overcome by adopting beamforming techniques at the cost of more hardware resources and higher power consumption. Shadowing effects are more difficult to cope as they are related to intrinsic microwave characteristics. Moreover, mmWave channels hold sparse nature with limited channel elements. Deploying UAV-assisted wireless networks can be an effective solution to mitigate this issue as it enables more line-of-sight (LoS) communications. In the near future, UAV-satellite communications \cite{UAVSAT} can enable more diverse Earth-space communication and hereby make 5G global access more robust and reliable.

As envisioned in \cite{ZENG2016}, UAV-aided wireless communications can fall into three representative categories of use cases, namely UAV-aided ubiquitous coverage, UAV-aided relaying, and UAV-aided information dissemination and data collection. The former two use cases are most likely applied to 5G base station (BS) offloading and wireless connectivity relaying. However, there are yet several major hurdles which prevent incorporating UAVs into 5G networks quickly and as smoothly as expected. First of all, UAVs could cause potential safety problems \cite{DING2018}; the Federal Aviation Administration (FAA) and many other countries' civil aviation authorities have regulated specific laws and rules on operating (flying) commercial UAVs \cite{FAA} with consideration for the weight, the maximum altitude and speed, the minimum distance from airports, constructions, vehicles, and people, etc. Enormous joint efforts from both policymakers and private industry are needed to achieve safe and efficient integration of UAVs into airspace and 5G networks. 

On the other hand, the UAV design is also confronted with various technical bottlenecks. One of the most significant challenges lies in limited onboard power. Take a 3-pound mainstream miniature UAVs (mini-UAV) for example, it is usually equipped with one 1-pound lithium polymer (LiPo) battery that is traditionally known to hold high power density. Although such a battery may contain a total energy of more than 80 watt-hours (Whs), it can barely support a maximum flight time of more than 30 minutes. Adding 5G wireless communication functions into UAVs will necessitate additional payload and additional power-hungry wireless hardware that further limits the operation duration. In addition, adverse weather conditions can pose more serious challenges to UAV operating time and 5G wireless communication quality.

In this article, we first present a brief review of worldwide research and development (R\&D) progress of UAV-enabled next-generation wireless communications, followed by an investigation on several main types of UAVs and the corresponding 5G application scenarios. Next, we propose a novel design of distributed and multi-layer UAVs (DAMU) 5G wireless network. We perform a thorough analysis of technical challenges of DAMU 5G network and feasible solutions for both wireless communication and power transfer. Finally, we present the analysis and modeling of atmospheric attenuation for the DAMU 5G network in typical meteorological conditions.

\section{UAV Enabled Wireless Communication and Power Transfer}
Widely recogonized as one of the biggest constraints of UAV-enabled wireless networks is the limited onboard energy. In a so-called 5G enabled UAV system, not only should the cellular communications system be accommodated, but also an extra control and non-payload communication (CNPC) system that should have outstanding performance in latency and security \cite{DING2018}. Generally, the aircraft (either fixed-wing or rotary-wing) engine consumes much higher energy than the communication systems of a microcell or picocell. Several solutions have been proposed to enhance the systematic robustness, the flight time, and the energy efficiency of a UAV system. 

The first type of solution is derived from the wireless system design perspective. In \cite{3D}, the authors presented algorithms to maximize the number of users by decoupling energy-efficient 3D placement of a UAV-BS within the region of interest (RoI). On the other hand, UAV path planning plays a critical role in obtaining a satisfactory energy efficiency and quality of service (QoS), and the planned paths depend on specific application scenarios. As an example, work in \cite{JIANG} optimized UAV’s flying direction for uplink communications by assuming a constant speed that the UAV maintains. In addition, the UAV trajectory optimization in \cite{ZENG2017} took into account the propulsion energy consumption of fixed-wing UAVs.

The second type of approach is to directly advance the energy resource technology and improve the energy management. In fact, from the energy density or specific energy (MJ/kg) perspective, both gasoline and jet fuel are at least 20 times higher than the LiPo battery widely used for mini-UAVs. In spite of the facts that the electric motor generally demonstrates much higher efficiency and speed adjustment capability than the petrol engines, the energy density gap cannot be filled up easily. As predicted, petrol engines or hybrid-electric engines should play a critical role in future UAVs used in 5G communications to achieve longer flight duration. 

New energy systems and energy harvesting techniques could further accelerate the pace of implementing 5G-oriented UAV networks. For example, \cite{SOLAR} presents a solar power management system (SPMS) for aircraft and UAV applications, with a maximum power tracking system (MPTS) to increase the operating efficiency of solar cells. Moreover, some other techniques for addressing energy transfer and storage challenges, such as wireless power transfer (WPT) and laser power beam techniques (developed and first demonstrated by Powerlight Technologies\footnote{\noindent https://powerlighttech.com/ [Accessed: 31-Mar-2018] }) can be also integrated to  possibly co-enable 24-hour flight working without landing or refueling. The authors in \cite{LASER} have proposed a throughput maximization scheme for balancing tradeoffs between laser energy harvesting and wireless communication performance. 

\begin{table}[h]
\caption{comparison of three categories of mainstream UAVs} \label{tab:UAV}
\newcommand{\tabincell}[2]{\begin{tabular}{@{}#1@{}}#2\end{tabular}}
 \centering
 \begin{threeparttable}
 \begin{tabular}{|c|c|c|c|c|c|c|c|}\hline

        \tabincell{c}{\textbf{Technology}}  &  \tabincell{c}{\textbf{Height(km)}} &  \tabincell{c}{ \textbf{Speed}}   &  \tabincell{c}{\textbf{Mobility and hovering}} & \tabincell{c}{\textbf{Energy resource} \\ \textbf{(primary first)} } & \tabincell{c}{\textbf{Endurance} } & \tabincell{c}{\textbf{Maximum} \\ \textbf{payload(kg)\tnote{4}}  } \\  
        \hline
        
        \tabincell{c}{Balloon\tnote{1} } &\tabincell{c}{Usually stays at \\the Stratosphere \\layer, $>$20 km} & \tabincell{c}{Slow} & \tabincell{c}{Low mobility,\\hovering supported} & \tabincell{c}{Solar cells,\\LiPo,\\petrol} & \tabincell{c}{Longest,\\from weeks to\\indefinite}  &\tabincell{c}{Large\tnote{5},\\$>$1000 kg} \\
        \hline
        
        \tabincell{c}{Fixed-wing\\UAV} &\tabincell{c}{Sea level-\\16 km} & \tabincell{c}{Fast\\(horizontally),\\medium\\(vertically)} & \tabincell{c}{Medium mobility,\\hovering not supported,\\minimum speed needs\\to be maintained} & \tabincell{c}{Petrol,\\solar cells,\\LiPo} & \tabincell{c}{Medium,\\from half day to\\days, or weeks\tnote{3}} & \tabincell{c}{Medium,\\$<$1000 kg} \\
        \hline

        \tabincell{c}{Rotary-wing\\UAV} &\tabincell{c}{Sea level-\\6 km} & \tabincell{c}{Medium\\(horizontally),\\fast\\(vertically)} & \tabincell{c}{Highest mobility\tnote{2},\\hovering supported} & \tabincell{c}{LiPo,\\petrol,\\solar cells} & \tabincell{c}{Low, less\\than 1 hour\\on LiPo\\battery.} & \tabincell{c}{Low,\\$<$100 kg} \\
        \hline

    \end{tabular}
   % \begin{tablenotes}
    %    \footnotesize
    %    \item[1]
    %  \end{tablenotes}
    
      \begin{tablenotes}
        \footnotesize
        \item[1] Can be Helium balloon, or Hydrogen balloon 
        \item[2] Electric motor facilitates highest mobility 
        \item[3] Solar powered fixed-wing UAVs can fly very long distance for weeks without landing
        \item[4] The maximum payload weight is for general scenarios, and it may vary in terms of specific UAV designs
        \item[5] Depends on balloon size, the altitude, temperature, wind speed, and atmospheric pressure
      \end{tablenotes}      
    
    \end{threeparttable}
\end{table}

\section{When Aerodynamics Meets 5G Communications}
At this section, we conduct investigation of UAVs communications from the aerodynamics perspective. As depicted in Fig. 1, in terms of the aerodynamics characteristics, there are three mainstream categories of UAVs: balloon, fixed-wing, and rotary-wing. A further comparison of these UAVs is summarized in Table I. Among them, balloons have been widely used for greater than 10 km, high altitude platforms (HAPs) and even ultra-high altitude (UHA) applications. For example, NASA's scientific balloons inflated with helium can lift heavy instruments (hundreds of kilograms) and stay at a height over 30 km for very long durations (intended for 100 days or longer). On the other hand, Google's Project Loon\footnote{\noindent https://x.company/loon/ [Accessed: 31-Mar-2018]} has successfully enabled a balloon network over 20 km high, extending the internet connectivity in rural and remote areas worldwide. In October 2017, Project Loon provided emergency long-term evolution (LTE) service recovery to Puerto Rico in the aftermath of Hurricane Maria. With solar panels and advanced predictive models of the winds and other metrological information from National Oceanographic and Atmospheric Administration (NOAA), the balloons can be navigated and deployed as requested. Furthermore, a balloon-UAV can facilitate quasi-stationary communications with propellers  adjusting balance and position.  

\begin{figure}
\centering
\subfigure [] {\includegraphics[scale=1.2]{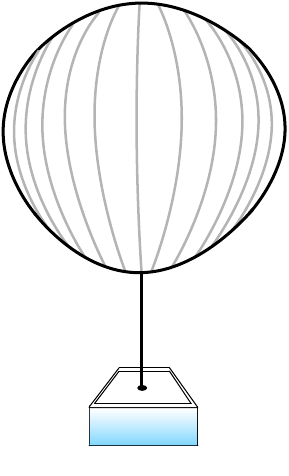}} \hspace*{ 0.5cm}\label{fig:FIG1A}
\subfigure [] {\includegraphics[scale=1.2]{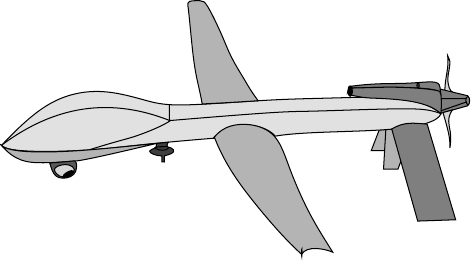}} \hspace*{ 0.5cm}\label{fig:FIG1B}
\subfigure [] {\includegraphics[scale=1.2]{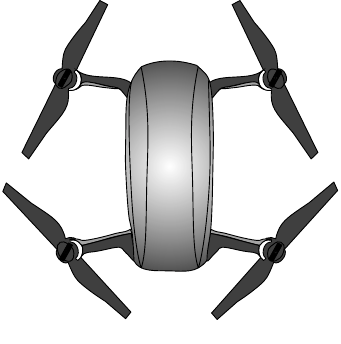}} \hspace*{ 0.5cm}\label{fig:FIG1C}
\caption{Three main categories of UAVs: (a) Balloon (b) Fixed-wing and (c) Rotary-wing (quadcopter).}\label{fig:FIG1}
\end{figure}

The second type of UAV is the fixed-wing UAV (FW-UAV); the most famous and successful example is General Atomics MQ-1 Predator first introduced in 1995. Normally, a FW-UAV can achieve a very wide range of altitude with the fastest horizontal speed due to powerful turbine engines. The maximum payload weight depends on the lift force that can be calculated using the following formula \cite{PILOT}:

\begin{equation}\label{eq:LIFT}
L=\frac{C_{\text{L}} \times \rho \times V^2 \times A}{2}
\end{equation}

\noindent where $L$ is the lift force which must equal the airplane's weight in pounds; $C_L$ is the coefficient of lift, which is determined by the airfoil type and angle of attack (AOA); $\rho$ is the air density and its specific value can be checked from the atmospheric model of the International Standard Atmosphere (ISA); $V$ stands for the velocity of the airfoil; and $A$ is the surface area. Therefore, the lift force is proportional to the wing area for the fixed-wing aircraft. Take a medium-sized FW-UAV (with 11 m$^2$ effective wing area) for instance, when flying with an AOA of 15 degrees, at 5 km high and 50 m/s (180 km/h), it can generate a lift force of 11800 Newtons or 1200 kg. However, if used for future 5G, a FW-UAV has to maintain a minimum speed to carry the weight of both equipment and UAV itself. On the other hand, no matter whether for functioning as a 5G aerial BS or relay, it is desirable for a FW-UAV to fly as slow as possible to minimize the Doppler effect and avoid complicating channel modeling and system design. Moreover, hybrid-electric engines and solar panels can further improve the energy efficiency and flight time to enable cost-effective UAV enabled 5G services. The solar-powered aircraft, Solar Impulse, is built with electric motors, lithium-ion (Li-ion) batteries and solar panels, and can realize very long-duration flights (118 hours) without landing. 

The third type of UAV is known as the rotary-wing UAV (RW-UAV) and has been popularly deployed in the consumer grade UAV market, particularly as a LiPo battery powered quadcopter as depicted in Fig.1(c). Such a RW-UAV can achieve very high aerodynamic flexibility and mobility with reliable hovering capability. In the UAV-based delivery system developed by Amazon Prime Air \footnote{https://www.amazon.com/Amazon-Prime-Air/b?ie=UTF8\&node=8037720011 [Accessed: 31-Mar-2018]}, RW-UAV with multiple propellers is demonstrated. However, the battery significantly limits its flight time and payload, consequently, the petrol engine based RW-UAV equipped with 6 or 8 propellers for industrial applications are developed for longer flight times (generally 3-5 times), better balance, and carrying greater payload. Nevertheless, the overall energy efficiency of RW-UAV is much lower than FW-UAV and balloon-UAV. Its application for 5G is duration and weight constrained due to the limited on-board energy until there is an effective solution to solve the energy puzzle. Besides the three major categories of UAVs, there are some hybrid aircraft designs, e.g., Bell Boeing V-22 Osprey, which take advantage of the virtues of all categories to achieve better design tradeoffs. 

\section{Distributed And Multi-Layer UAV Networking}
\begin{figure}[!t]
\begin{center}
\includegraphics[width=5in]{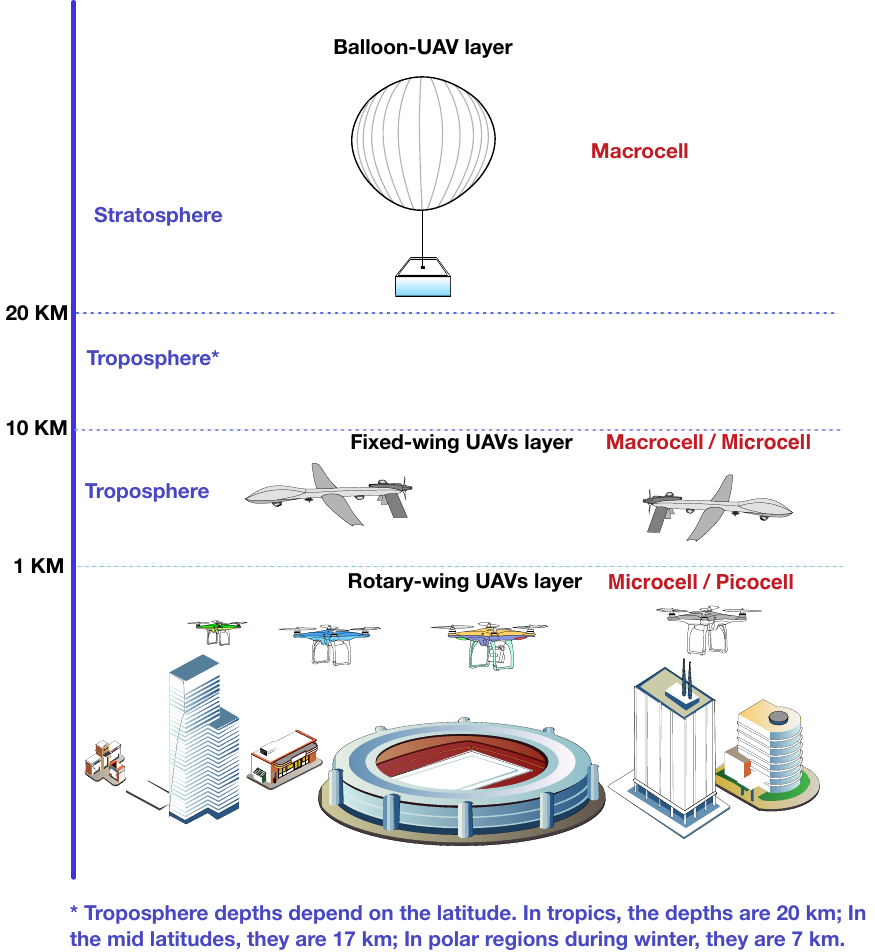}
\caption{Distributed and multi-layer UAVs (DAMU) network architecture.}\label{FIG2}
\end{center}
\end{figure}

From the above investigation, a balloon-UAV may be the most suitable aircraft for carrying heavy 5G infrastructure equipment and hovering over the sky with the longest duration. Considering the significant height and coverage (with a radius more than 20 km) it can achieve, an energy-effective balloon-UAV can serve as a 5G powerful macrocell base station that could weigh up to hundreds of kilograms. On the other hand, a FW-UAV may carry a 5G macrocell/microcell and fly within a flexible altitude below 10 km. Moreover, a RW-UAV is more ideal for installing lightweight 5G equipment (such as a picocell) and executing limited duration tasks that require fast deployment. Based on the characteristics of different UAVs, we propose a distributed and multi-layer UAV (DAMU) network architecture as depicted in Fig.~\ref{FIG2}. 

\begin{enumerate}

\item The balloon-UAV functions as a quasi-stationary cellular tower and generally stays at the stratosphere layer, most likely above 20 km. In theory, the balloon can be recycled and relaunched before or after a regular maintenance. 

\item The fixed-wing UAVs are generally deployed below 10 km and above 1 km. In order to minimize the Doppler shift and the associated system design challenges, FW-UAVs need to cruise at the slowest speed possible. 

\item The rotary-wing UAVs are normally dispatched below 1 km, serving as microcell/picocell base stations. Low cruise altitudes can enable RW-UAVs to be frequently and quickly recharged or replenished. 

\end{enumerate}

\section{DAMU Wireless Communications And Power Transfer}

\begin{figure}[!t]
\begin{center}
\includegraphics[width=6.5in]{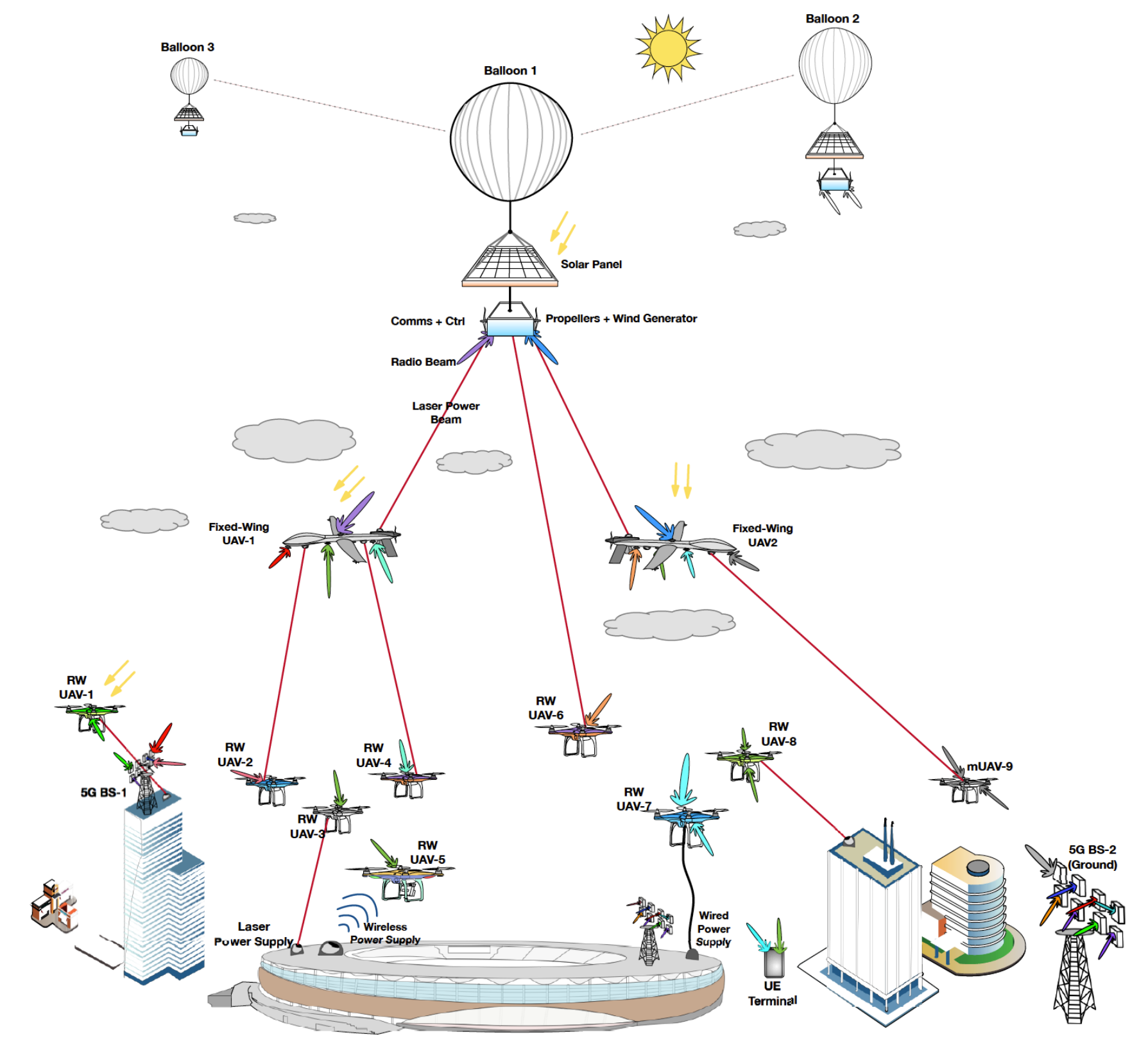}
\caption{Wireless communications and power transfer in a DAMU enabled 5G wireless communication and power transfer network.}\label{FIG3}
\end{center}
\end{figure}

Based on the DAMU network architecture, a comprehensive solution for 5G wireless communications and power transfer heterogeneous networks is hereby proposed and depicted in Fig.~\ref{FIG3}.

\begin{enumerate}

\item The solar-powered balloon-UAVs are equipped with solar panels, batteries, and wind generators to realize a self sustaining energy system. Staying at stratosphere layer facilitates efficient solar energy harvesting as no weather occurs. 5G macrocell communications and CNPC systems are integrated and enabled in the ballon-UAVs. The balloons provide both 5G new radio (NR) facility and backwards compatibility to legacy 3GPP standards; in addition, the balloons communicate with other aerial or ground base stations, as well as ground terminals. In a typical line-of-sight (LoS) communication scenario for 5G mmWave bands, Balloon 1 operates its phased arrays to form multiple beams to align with the beams from FW-UAV-1 and FW-UAV-2, respectively. 

\item Fixed-wing UAVs serve as either a macrocell or a microcell. In a representative 5G usage scenario, a FW-UAV may communicate with both a balloon-UAV macrocell and multiple RW-UAVs based picocells. As depicted in Fig.~\ref{FIG3}, RW-UAV-5 is live-streaming a sport event over the stadium and uploading the ultra-high resolution (UHD) video to RW-UAV-3 that further communicates with FW-UAV-1 using beamforming. A FW-UAV can generate multiple mmWave beams to increase spatial multiplexing gain and channe capacity, and mitigate the inteference as well, by adopting a distributed phased array MIMO (DPA-MIMO) reconfigruable architecture in \cite{HUO}.

\item Rotary-wing UAVs are dispatched and deployed mainly for microcell/picocell applications below an altitude of 1 km. They can enable fast 5G access and services whenever or wherever there is such a need. A RW-UAV can be connected to either aerial/ground base stations or aerial/ground user terminals. 

\begin{figure}[!t]
\begin{center}
\includegraphics[width=4in]{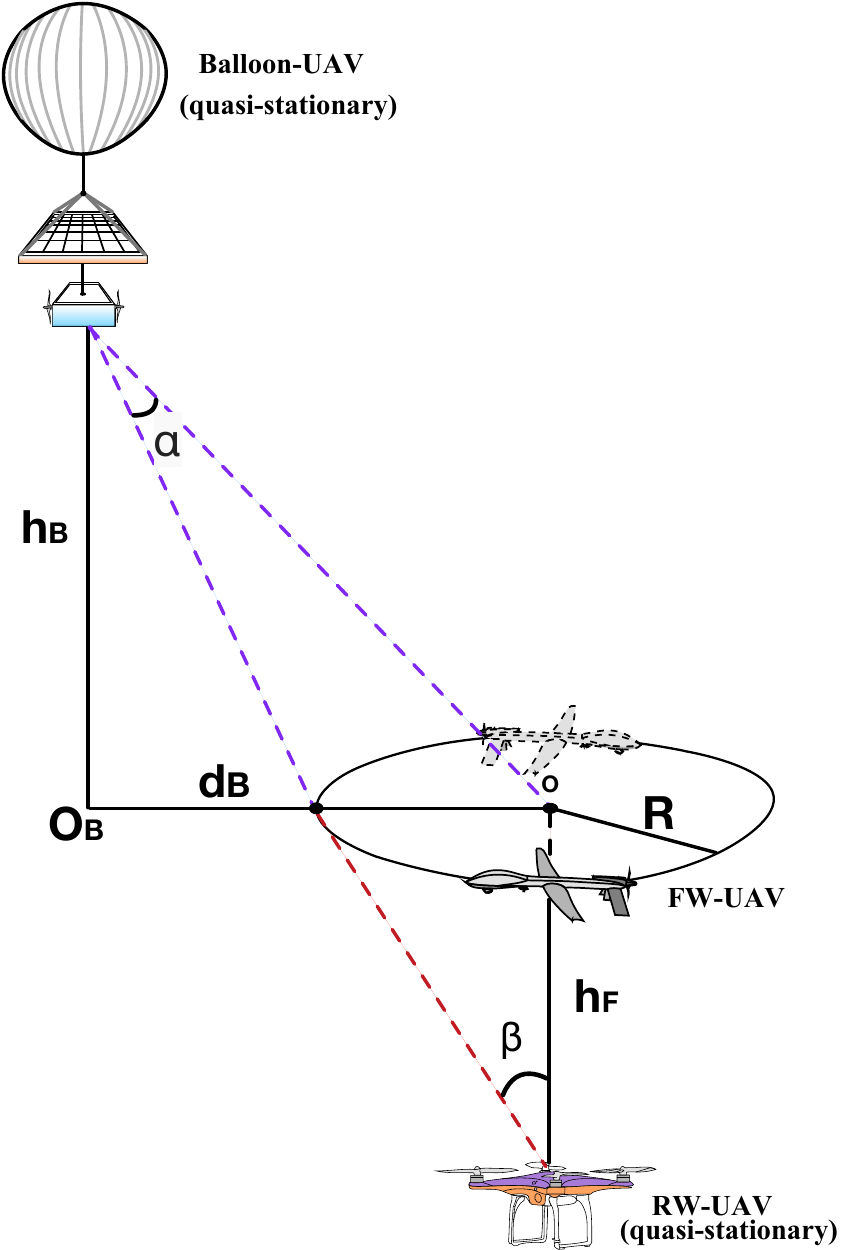}
\caption{FW-UAV circular cruising mode in a DAMU system.}\label{FIG4}
\end{center}
\end{figure}

\item Under specific conditions, each layer of UAVs should be able to work independently to sustain a full-function 5G network when other layers of UAVs are not available. 

\item When a FW-UAV based aerial BS enables 5G communications, its velocity should be well maintained at the lowest possible speed, and its flight path needs to be well planned. Assume that a FW-UAV is needed to provide 5G service coverage for some area of interest with a radius of 2 km, we need to dispatch it to fly at least 2 km high. In order to build a reliable and easy-approaching communication tunnel, we first program the FW-UAV fly with a (pre-defined) specific pattern that is known and easy to follow by other aerial/ground BSs and terminals, at its minimum speed. Moreover, this pattern needs to enable good energy efficiency from aerodynamics perspective. For example, the cruising path can be a simple circle with a small radius of turn which can be calculated as \cite{PILOT}

\begin{equation}\label{eq:RADIUS}
R=\frac{V^2}{11.26 \times \tan(\theta)}
\end{equation}

where $R$ is in the unit of feet, $V$ is the velocity in the unit of knot, and $\theta$ is the bank angle. According to emerging medium-sized FW-UAV ($>$ 100 kg of extra payload) designs, a minimum velocity can be maintained at 10 m/s (19.4 knot) at the sea level.

Demonstrating circular cruising as depicted in Fig. ~\ref{FIG4}, we assume the FW-UAV flies 2 km high with a speed of 20 m/s (38.8 knot). If the bank angle is 30$^{\circ}$, the radius of turn, $R$, is calculated to be 232.4 feet or 70.8 meters. 
Furthermore, assume a balloon-UAV that hovers 20 km high is 5 km horizontally away from the center of circle, while a quasi-stationary RW-UAV is precisely 1.9 km vertically beneath the center of circle, or 100 m above the ground level. $\alpha$ and $\beta$ are calculated to be 0.42$^{\circ}$ and 2.13$^{\circ}$, respectively. In other words, the proposed circular cruising can conditionally minimize the spatial angle. This scheme leads to several benefits; first, pre-defined cruise path mitigates the localization; second, the challenges of mmWave beamforming, beam tracking, and beam alignment are significantly mitigated, between FW-UAV and other BSs or terminals. 

\item In order to enable longer flight time, FW-UAVs and RW-UAVs may adopt solar energy harvesting and laser power transfer strategies. As depicted in Fig.~\ref{FIG3}, high power is beamed to FW-UAV-1 and FW-UAV-2 from Balloon 1 which may have harvested and accumulated significant solar energy. It is envisioned that the laser beam power transfer technique can address emergent airborne recharging requests for the 5G aerial base stations. However, one of the prominent challenges of laser power transfer stems from the atmospheric losses that depend on the visibility and weather conditions. Take 1550 nm laser for instance, during a clear day with 23 km visibility, the loss is only 0.2 dB/km, whereas the loss will be up to 4 dB/km when encountering haze weather \cite{LASER2}. Moreover, a heavy fog with 0.05 km visibility can result in a loss as large as 272 dB/km. Subsequently, using laser power transfer is, to a large extent, weather and distance dependent. Additionally, laser safety is another major concern of the DAMU network design since the laser power can be up to hundreds of watts. Machine-learning aided computer vision and sensor techniques, as well as distributed laser beam charging may help achieve safe laser energy harvesting.   

\item Some other alternative power supply methods are also critical complements of a DAMU network. For example, as shown in Fig.~\ref{FIG3}, RW-UAV-5 is charged through conventional microwave when it is not far apart from the charger. Alternatively, RW-UAV-3 is wired through a power cable to a power source, and the cable length can be more than 100 meters according to experimental practice.    

\end{enumerate}

\section{UAV Communications And Meteorological Conditions}

For UAV communications, air-to-ground (A2G) channel characteristics are significantly different from terrestrial ones. In order of blockage, there are mainly three classes of links; namely line-of-sight (LoS), obstructed line-of-sight (OLoS), and non-line-of-sight (NLoS). In A2G communications, the probability of occurrence (LoS or NLoS) is a function of environments and the channel modeling has been thoroughly reviewed in \cite{TUTORIAL}. In our proposed DAMU network, air-to-air (A2A) communications among different layers and UAVs above ground level are the application scenarios of interest. 

As previously discussed, weather conditions play a critical role in DAMU 5G networks by having apparent impact on wireless communications, power transfer, and UAV working status. It should be noted that global weather and climate patterns are dramatically diverse. Therefore, in this section, we focus on A2A attenuation modeling for frequency below 100 GHz with weather factors taken into account.  

First, atmospheric humidity largely affects gaseous attenuation, particularly at mmWave bands and above. The quantitative analysis of gaseous attenuation over frequency as a variable of water vapor density, is given in prediction models recommended by ITU \cite{ITU2017}. Generally, gaseous attenuation due to water vapor increases over frequency. At sea level and under standard atmosphere (7.5 g/m$^3$ water vapor density), the total gaseous attenuation (dry air plus water vapor) grows to more than 1 dB/km from 53 to 67 GHz and sharply peaks at 15 dB/km for 61 GHz. This is attributed, to a large extent, by oxygen absorption.   

From the meteorological perspective, precipitation can be in the forms of drizzle, rain, sleet, snow, and hail. According to ITU rain attenuation models, heavy rainfall can cause significant attenuation at 5G mmWave bands. When the weather gets cloudy and foggy, there are two methodologies to calculate attenuation \cite{ITU2017}. The first one is to obtain the specific attenuation (dB/km) within a cloud or fog, which can be written as

\begin{equation}\label{eq:ATTENUATION}
\gamma_c=K_lM
\end{equation}

\noindent where $K_l$ is the specific attenuation coefficient ((dB/km)/(g/m$^3$)), and it is a function about frequency and dielectric permittivity of water; $M$ is the liquid water density (LWD) in cloud or fog (g/m$^3$). For medium and thick fog, the LWD is around 0.05 g/m$^3$ and 0.5 g/m$^3$, respectively \cite{CLOUD}. Additionally, advection fog can be vertically thick more than 2 km above ground level.  

Furthermore, the second methodology is to calculate attenuation due to clouds for a given probability. This attenuation is correlated with the statistics of the total columnar content of cloud liquid water L (kg/m$^2$) for a given geographic location, and it is expressed as

\begin{equation}\label{eq:ATTENUATION2}
A=\frac{LK_l}{\sin(\theta)}
\end{equation}

\noindent where $K_l$ is the specific attenuation coefficient, and $\theta$ is the elevation angle within a range from 5 to 90 degrees. The value $L$ can be checked from map based data files \cite{ITU2017}. For example in some regions of Southeast Asia, $L$ can be as high as 2 for a yearly exceedance probability of 1\%. However, the second method is not able to directly estimate the worst-case A2A communications scenarios, especially before heavy rainfall.

Out of many forms and types of clouds in the Earth’s atmosphere, cumulonimbus clouds are a dense, towering vertical clouds that can be very tall and thick with the highest LWD. According to cloud thickness estimation from GOES-8 satellite data \cite{CLOUD}, precipitating cumulus clouds (precursor of cumulonimbus cloud) have a mean thickness of 9.32 km. Using these variables, we can plot the atmospheric attenuation over frequency, with various meteorological conditions introduced.

\begin{figure}[!t]
\begin{center}
\includegraphics[width=5in]{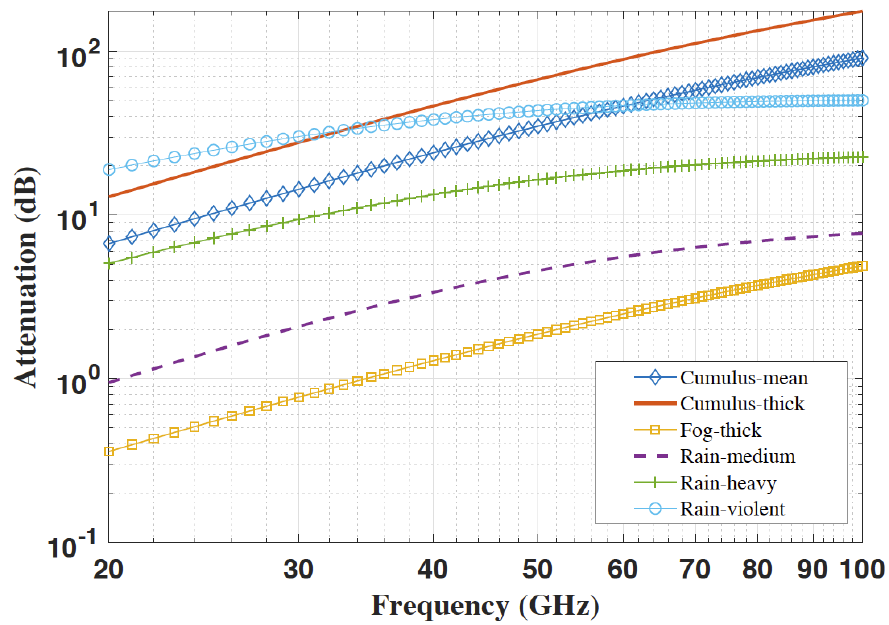}
\caption{Attenuation due to typical meteorological conditions for 5G high frequency bands.}\label{FIG5}
\end{center}
\end{figure}

Assuming the elevation angle between a balloon-UAV (at 20 km) and FW-UAV (hovering at 1 km) is 90 degrees, a LoS channel normally exists during a clear day. As depicted in Fig.~\ref{FIG5}, if a very thick cumulonimbus cloud with high LWD (12 km, 3 g/m$^3$) exists, it causes the highest attenuation for frequencies over 40 GHz. Moreover, if a 2 km vertically thick advection fog emerges, the resulting attenuation is 0.68 and 1.28 dB at 28 GHz and 40 GHz, respectively. If the precipitation happens, the attenuation caused by medium rain, heavy rain, and violent rain dramatically varies. For example, a violent rain (100 mm/hour) can result in a 38.3 dB attenuation at 40 GHz, compared to a 3.4 dB attenuation caused by medium rain. Therefore, a dense cumulonimbus cloud before or during precipitation will lead to the most significant attenuation. The effects of weather should be carefully and thoroughly considered with other types of propagation loss when conducting the DAMU 5G network design and link budget calculation.  

\section{Conclusion}
UAV enabled 5G communications is promising to become an immediate reality bringing both opportunities and challenges. In this article, investigations were first conducted of the most recent R\&D progress of UAVs and UAV-enabled wireless communication technologies. Facts were unveiled that the design challenges of UAV enabled 5G communications are more pressing than terrestrial 5G communications. This is mainly because a UAV enabled 5G system necessitates more comprehensive system designs by considering several major critical factors, namely wireless communication system, aerodynamic constraints, and meteorological variables. Next, a novel distributed and multi-layer UAVs (DAMU) networking architecture was presented for 5G wireless communication and beyond. The DAMU networking architecture is hierarchical, flexible, and can be reconfigured in terms of specific deployment and application scenarios. In addition, we have taken into account several key technical enablers such as the practical aerodynamic design rules for different types of UAV designs, energy harvesting techniques, and power transfer techniques. Furthermore, we have conducted numerical analysis of the attenuation introduced by typical meteorological conditions to give the guidelines for an overall link budget calculation and system robustness design analysis.

%\section*{Acknowledgments}


\begin{thebibliography}{1}

\bibitem{ITU2015}
Rec. ITU-R M.2083-0, ``IMT Vision - Framework and overall objectives of the future development of IMT for 2020 and beyond," Sep. 2015.

\bibitem{UAVSAT}
J. Zhao, F. Gao, Q. Wu, S. Jin, Y. Wu, and W. Jia, ``Beam tracking for UAV mounted SatCom on-the-move with massive antenna array," \emph{IEEE J. Sel. Areas Commun.}, vol. 36, no. 2, pp. 363--375, Feb. 2018.

\bibitem{ZENG2016}
Y. Zeng, R. Zhang, and T. J. Lim, ``Wireless communications with unmanned aerial vehicles: Opportunities and challenges," \emph{IEEE Commun. Mag.}, vol. 52, no. 2, pp. 122--130, Feb. 2014.

\bibitem{DING2018}
S. F. Yunas, M. Valkama and J. Niemela, ``Spectral and energy efficiency of ultra-dense networks under different deployment strategies," \emph{IEEE Commun. Mag.}, Vol. 53, Vol. 1, pp. 90--100, Jan. 2015.

\bibitem{FAA}
Federal Aviation Administration, ``Unmanned aircraft systems,'' [Online]. Available: https://www.faa.gov/uas/

\bibitem{3D}
M. Alzenad, A. El-keyi, F. Lagum, and H. Yanikomeroglu, ``3D placement of an unmanned aerial vehicle base station (UAV-BS) for energy-efficient maximal coverage,'' \emph{IEEE Wireless Commun. Lett.}, vol. 6, no. 4, pp. 434--437, Aug. 2017.

\bibitem{JIANG}
F. Jiang and A. L. Swindlehurst, ``Optimization of UAV heading for the ground-to-air uplink,'' \emph{IEEE J. Sel. Areas Commun.}, vol. 30, no. 5, pp. 993--1005, Jun. 2012.

\bibitem{ZENG2017}
Y. Zeng and R. Zhang, ``Energy-efficient UAV communication with trajectory optimization,'' \emph{IEEE Trans. Wireless Commun.}, vol. 16, no. 6, pp. 3747--3760, Jun. 2017.

\bibitem{SOLAR}
J. K. Shiau, D. M. Ma, P. Y. Yang, G. F. Wang, and J. H. Gong, ``Design of a solar power management system for an experimental UAV,'' \emph{IEEE Trans. Aerosp. Electron. Syst.}, vol. 45, no. 4, pp. 1350--1360, Oct. 2009.

\bibitem{LASER}
J. Ouyang, Y. Che, J. Xu, and K. Wu, ``Throughput maximization for laser-powered UAV wireless communication systems,'' to appear in \emph{IEEE ICC 2018}, [Online]. Available: https://arxiv.org/abs/1803.00690

\bibitem{PILOT}
Federal Aviation Administration, ``Pilot's handbook of aeronautical knowledge: FAA-H-8083-25B,'' Newcastle, WA, USA: Aviation Supplies and Academics, Inc., 2016. 

\bibitem{HUO}
Y. Huo, X. Dong, and W. Xu, ``5G Cellular User Equipment: From Theory to Practical Hardware Design,'' \emph{IEEE Access}, vol. 5, pp. 13992--14010, 2017.

\bibitem{LASER2}
I. I. Kim, B. McArthur, and E. Korevaar, ``Comparison of laser beam propagation at 785 nm and 1550 nm in fog and haze for optical wireless communications,'' in \emph{Proc. SPIE, Opt. Wireless Commun. III}, Boston, MA, USA, Nov. 2001, vol. 4214, pp. 26--37.

\bibitem{TUTORIAL}
M. Mozaffari, W. Saad, M. Bennis, Y.-H. Nam, and M. Debbah, ``A tutorial on UAVs for wireless networks: Applications, challenges, and open problems,'' [Online]. Available: https://arxiv.org/abs/1803.00680

\bibitem{ITU2017}
Recommendation ITU--R P.840-7, ``Attenuation due to clouds and fog,'' 2017.

\bibitem{CLOUD}
V. Chakrapani, D. R. Doelling, A. D. Rapp, and P. Minnis, ``Cloud thickness estimation from GOES-8 satellite data over the ARM-SGP site,'' in \emph{Proc. 12th ARM Science Team Meeting}, St. Petersburg, FL, U.S.A., Apr. 2002, pp. 1--7.

\end{thebibliography}
\end{document}